\documentclass[article,fullpage]{elsarticle}

\graphicspath{ {pix/} }

\usepackage{graphicx}% Include figure files
\usepackage{psfrag}%
\usepackage{xspace}%
\usepackage{dcolumn}% Align table columns on decimal point
\usepackage{bm}% bold math
\usepackage{amssymb}%
\usepackage{enumitem}%

\usepackage{lineno}% Enable numbering of text and display math
\makeatletter
\def\ps@pprintTitle{Published in PLB as DOI: 10.1016/j.physletb.2016.09.039
 \let\@oddhead\@empty
 \let\@evenhead\@empty
 \def\@oddfoot{}%
 \let\@evenfoot\@oddfoot}
\makeatother

\makeatletter
\g@addto@macro\endfrontmatter{\enlargethispage{-2\baselineskip}}
\makeatother

\journal{Physics Letters B}

\newcommand{\meanlnA}{\ensuremath{\langle\ln A\rangle}}
\newcommand{\eposlhc}{EPOS-LHC}
\newcommand{\lnA}{\ensuremath{\ln A}}

\newcommand{\xmaxs}{\ensuremath{X_\mathrm{max}}}
\newcommand{\smill}{\ensuremath{S(1000)}}
\newcommand{\smilla}{\ensuremath{S^*_{38}}}
\newcommand{\rGs}{\ensuremath{r_\mathrm{G}}}

\newcommand{\logenr}[2]{\ensuremath{\lg(E/\mathrm{eV})=#1-#2}}
\newcommand{\rGsmilla}{\ensuremath{r_\mathrm{G}(\xmax,\,\smilla)}}

\newcommand{\xmax}{\ensuremath{X_\mathrm{max}^{*}}}

\newcommand{\gsm}{g\,cm${}^{-2}$}

\begin{document}

\begin{frontmatter}

\title{Evidence for a mixed mass composition at the `ankle' in the
  cosmic-ray spectrum}
%

%\author{\input{2016-05-author_list_latex_authors.tex}}
%\address{\input{2016-05-author_list_latex_institutions.tex}}

\author{% created on 2016-06-15

A.~Aab$^{37}$,
P.~Abreu$^{70}$,
M.~Aglietta$^{48,47}$,
E.J.~Ahn$^{85}$,
I.~Al Samarai$^{29}$,
I.F.M.~Albuquerque$^{16}$,
I.~Allekotte$^{1}$,
P.~Allison$^{90}$,
A.~Almela$^{8,11}$,
J.~Alvarez Castillo$^{62}$
% ...
}
\address{% created on 2016-06-15

% needs \usepackage{enumitem}
\begin{description}[labelsep=0.2em,align=right,labelwidth=0.7em,labelindent=0em,leftmargin=2em,noitemsep]
\item[$^{1}$] Centro At\'omico Bariloche and Instituto Balseiro (CNEA-UNCuyo-CONICET), Argentina
\item[$^{2}$] Centro de Investigaciones en L\'aseres y Aplicaciones, CITEDEF and CONICET, Argentina
\item[$^{3}$] Departamento de F\'\i{}sica and Departamento de Ciencias de la Atm\'osfera y los Oc\'eanos, FCEyN, Universidad de Buenos Aires, Argentina
\item[$^{4}$] IFLP, Universidad Nacional de La Plata and CONICET, Argentina
\item[$^{5}$] Instituto de Astronom\'\i{}a y F\'\i{}sica del Espacio (IAFE, CONICET-UBA), Argentina
\item[$^{6}$] Instituto de F\'\i{}sica de Rosario (IFIR) -- CONICET/U.N.R.\ and Facultad de Ciencias Bioqu\'\i{}micas y Farmac\'euticas U.N.R., Argentina
\item[$^{7}$] Instituto de Tecnolog\'\i{}as en Detecci\'on y Astropart\'\i{}culas (CNEA, CONICET, UNSAM) and Universidad Tecnol\'ogica Nacional -- Facultad Regional Mendoza (CONICET/CNEA), Argentina
\item[$^{8}$] Instituto de Tecnolog\'\i{}as en Detecci\'on y Astropart\'\i{}culas (CNEA, CONICET, UNSAM), Centro At\'omico Constituyentes, Comisi\'on Nacional de Energ\'\i{}a At\'omica, Argentina
\item[$^{9}$] Observatorio Pierre Auger, Argentina
\item[$^{10}$] Observatorio Pierre Auger and Comisi\'on Nacional de Energ\'\i{}a At\'omica, Argentina
\item[$^{11}$] Universidad Tecnol\'ogica Nacional -- Facultad Regional Buenos Aires, Argentina
\item[$^{12}$] University of Adelaide, Australia
\item[$^{13}$] Centro Brasileiro de Pesquisas Fisicas (CBPF), Brazil
\item[$^{14}$] Universidade de S\~ao Paulo, Escola de Engenharia de Lorena, Brazil
\item[$^{15}$] Universidade de S\~ao Paulo, Inst.\ de F\'\i{}sica de S\~ao Carlos, S\~ao Carlos, Brazil
\item[$^{16}$] Universidade de S\~ao Paulo, Inst.\ de F\'\i{}sica, S\~ao Paulo, Brazil
\item[$^{17}$] Universidade Estadual de Campinas (UNICAMP), Brazil
\item[$^{18}$] Universidade Estadual de Feira de Santana (UEFS), Brazil
\item[$^{19}$] Universidade Federal de Pelotas, Brazil
\item[$^{20}$] Universidade Federal do ABC (UFABC), Brazil
\item[$^{21}$] Universidade Federal do Paran\'a, Setor Palotina, Brazil
\item[$^{22}$] Universidade Federal do Rio de Janeiro (UFRJ), Instituto de F\'\i{}sica, Brazil
\item[$^{23}$] Universidade Federal Fluminense, Brazil
\item[$^{24}$] Universidad Industrial de Santander, Colombia
\item[$^{25}$] Institute of Physics (FZU) of the Academy of Sciences of the Czech Republic, Czech Republic
\item[$^{26}$] Palacky University, RCPTM, Czech Republic
\item[$^{27}$] University Prague, Institute of Particle and Nuclear Physics, Czech Republic
\item[$^{28}$] Institut de Physique Nucl\'eaire d'Orsay (IPNO), Universit\'e Paris 11, CNRS-IN2P3, France
\item[$^{29}$] Laboratoire de Physique Nucl\'eaire et de Hautes Energies (LPNHE), Universit\'es Paris 6 et Paris 7, CNRS-IN2P3, France
\item[$^{30}$] Laboratoire de Physique Subatomique et de Cosmologie (LPSC), Universit\'e Grenoble-Alpes, CNRS/IN2P3, France
\item[$^{31}$] Bergische Universit\"at Wuppertal, Department of Physics, Germany
\item[$^{32}$] Karlsruhe Institute of Technology, Institut f\"ur Experimentelle Kernphysik (IEKP), Germany
\item[$^{33}$] Karlsruhe Institute of Technology, Institut f\"ur Kernphysik (IKP), Germany
\item[$^{34}$] Karlsruhe Institute of Technology, Institut f\"ur Prozessdatenverarbeitung und Elektronik (IPE), Germany
\item[$^{35}$] RWTH Aachen University, III.\ Physikalisches Institut A, Germany
\item[$^{36}$] Universit\"at Hamburg, II.\ Institut f\"ur Theoretische Physik, Germany
\item[$^{37}$] Universit\"at Siegen, Fachbereich 7 Physik -- Experimentelle Teilchenphysik, Germany
\item[$^{38}$] Gran Sasso Science Institute (INFN), L'Aquila, Italy
\item[$^{39}$] INAF -- Istituto di Astrofisica Spaziale e Fisica Cosmica di Palermo, Italy
\item[$^{40}$] INFN Laboratori Nazionali del Gran Sasso, Italy
\item[$^{41}$] INFN, Gruppo Collegato dell'Aquila, Italy
\item[$^{42}$] INFN, Sezione di Catania, Italy
\item[$^{43}$] INFN, Sezione di Lecce, Italy
\item[$^{44}$] INFN, Sezione di Milano, Italy
\item[$^{45}$] INFN, Sezione di Napoli, Italy
\item[$^{46}$] INFN, Sezione di Roma ``Tor Vergata``, Italy
\item[$^{47}$] INFN, Sezione di Torino, Italy
\item[$^{48}$] Osservatorio Astrofisico di Torino (INAF), Torino, Italy
\item[$^{49}$] Universit\`a del Salento, Dipartimento di Ingegneria, Italy
\item[$^{50}$] Universit\`a del Salento, Dipartimento di Matematica e Fisica ``E.\ De Giorgi'', Italy
\item[$^{51}$] Universit\`a dell'Aquila, Dipartimento di Scienze Fisiche e Chimiche, Italy
\item[$^{52}$] Universit\`a di Catania, Dipartimento di Fisica e Astronomia, Italy
\item[$^{53}$] Universit\`a di Milano, Dipartimento di Fisica, Italy
\item[$^{54}$] Universit\`a di Napoli ``Federico II``, Dipartimento di Fisica ``Ettore Pancini``, Italy
\item[$^{55}$] Universit\`a di Roma ``Tor Vergata'', Dipartimento di Fisica, Italy
\item[$^{56}$] Universit\`a Torino, Dipartimento di Fisica, Italy
\item[$^{57}$] Benem\'erita Universidad Aut\'onoma de Puebla (BUAP), M\'exico
\item[$^{58}$] Centro de Investigaci\'on y de Estudios Avanzados del IPN (CINVESTAV), M\'exico
\item[$^{59}$] Unidad Profesional Interdisciplinaria en Ingenier\'\i{}a y Tecnolog\'\i{}as Avanzadas del Instituto Polit\'ecnico Nacional (UPIITA-IPN), M\'exico
\item[$^{60}$] Universidad Aut\'onoma de Chiapas, M\'exico
\item[$^{61}$] Universidad Michoacana de San Nicol\'as de Hidalgo, M\'exico
\item[$^{62}$] Universidad Nacional Aut\'onoma de M\'exico, M\'exico
\item[$^{63}$] Institute for Mathematics, Astrophysics and Particle Physics (IMAPP), Radboud Universiteit, Nijmegen, Netherlands
\item[$^{64}$] KVI -- Center for Advanced Radiation Technology, University of Groningen, Netherlands
\item[$^{65}$] Nationaal Instituut voor Kernfysica en Hoge Energie Fysica (NIKHEF), Netherlands
\item[$^{66}$] Stichting Astronomisch Onderzoek in Nederland (ASTRON), Dwingeloo, Netherlands
\item[$^{67}$] Institute of Nuclear Physics PAN, Poland
\item[$^{68}$] University of \L{}\'od\'z, Faculty of Astrophysics, Poland
\item[$^{69}$] University of \L{}\'od\'z, Faculty of High-Energy Astrophysics, Poland
\item[$^{70}$] Laborat\'orio de Instrumenta\c{c}\~ao e F\'\i{}sica Experimental de Part\'\i{}culas -- LIP and Instituto Superior T\'ecnico -- IST, Universidade de Lisboa -- UL, Portugal
\item[$^{71}$] ``Horia Hulubei'' National Institute for Physics and Nuclear Engineering, Romania
\item[$^{72}$] Institute of Space Science, Romania
\item[$^{73}$] University of Bucharest, Physics Department, Romania
\item[$^{74}$] University Politehnica of Bucharest, Romania
\item[$^{75}$] Experimental Particle Physics Department, J.\ Stefan Institute, Slovenia
\item[$^{76}$] Laboratory for Astroparticle Physics, University of Nova Gorica, Slovenia
\item[$^{77}$] Universidad Complutense de Madrid, Spain
\item[$^{78}$] Universidad de Alcal\'a de Henares, Spain
\item[$^{79}$] Universidad de Granada and C.A.F.P.E., Spain
\item[$^{80}$] Universidad de Santiago de Compostela, Spain
\item[$^{81}$] Case Western Reserve University, USA
\item[$^{82}$] Colorado School of Mines, USA
\item[$^{83}$] Colorado State University, USA
\item[$^{84}$] Department of Physics and Astronomy, Lehman College, City University of New York, USA
\item[$^{85}$] Fermi National Accelerator Laboratory, USA
\item[$^{86}$] Louisiana State University, USA
\item[$^{87}$] Michigan Technological University, USA
\item[$^{88}$] New York University, USA
\item[$^{89}$] Northeastern University, USA
\item[$^{90}$] Ohio State University, USA
\item[$^{91}$] Pennsylvania State University, USA
\item[$^{92}$] University of Chicago, USA
\item[$^{93}$] University of Hawaii, USA
\item[$^{94}$] University of Nebraska, USA
\item[$^{95}$] University of New Mexico, USA
\item[$^{a}$] School of Physics and Astronomy, University of Leeds, Leeds, United Kingdom
\item[$^{b}$] Max-Planck-Institut f\"ur Radioastronomie, Bonn, Germany
\item[$^{c}$] also at Vrije Universiteit Brussels, Brussels, Belgium
\item[$^{d}$] now at Deutsches Elektronen-Synchrotron (DESY), Zeuthen, Germany
\item[$^{e}$] SUBATECH, \'Ecole des Mines de Nantes, CNRS-IN2P3, Universit\'e de Nantes
\item[$^{f}$] Los Alamos National Laboratory, USA
\end{description}
}

\date{\today}

\begin{abstract}
%\begin{linenumbers}

We report a first measurement for ultra-high energy cosmic rays of the
correlation between the depth of shower maximum and the signal in the
water Cherenkov stations of air-showers registered simultaneously by
the fluorescence and the surface detectors of the Pierre Auger
Observatory. Such a correlation measurement is a unique feature of a
hybrid air-shower observatory with sensitivity to both the
electromagnetic and muonic components. It allows an accurate
determination of the spread of primary masses in the cosmic-ray flux.
Up till now, constraints on the spread of primary masses have been
dominated by systematic uncertainties. The present correlation
measurement is not affected by systematics in the measurement of the
depth of shower maximum or the signal in the water Cherenkov
stations. The analysis relies on general characteristics of air
showers and is thus robust also with respect to uncertainties in
hadronic event generators.  The observed correlation in the energy
range around the `ankle' at \logenr{18.5}{19.0} differs significantly
from expectations for pure primary cosmic-ray compositions. A light
composition made up of proton and helium only is equally inconsistent
with observations.  The data are explained well by a mixed composition
including nuclei with mass $A > 4$.  Scenarios such as the proton dip
model, with almost pure compositions, are thus disfavoured as the sole
explanation of the ultrahigh-energy cosmic-ray flux at Earth.
%\end{linenumbers}
\end{abstract}

%pacs{96.50.sd, 98.70.Sa, 13.85.Tp}


\end{frontmatter}

%\linenumbers
%%%%%%%%%%%%%%%%%
%% Introduction.
%%%%%%%%%%%%%%%%%

\section{Introduction}

An important quantity to characterize the composition of cosmic rays
is the spread in the range of masses in the primary beam. In
theoretical source models regarding protons as the dominant particle
type, the composition is expected to be (almost) pure, while in other
scenarios also allowing heavier nuclei to be accelerated, a mixed
composition is predicted.  For instance, in the `dip'
model~\cite{berezinsky_dip_prl2005,berezinsky_dip_ankle_2006}, two
observed features of the energy spectrum could be naturally understood
as a signature of proton interactions during propagation (ankle at
$\lg(E/\mathrm{eV}) \simeq 18.7$ from pair-production and flux
suppression at $\lg(E/\mathrm{eV}) \simeq 19.6$ from photopion
production). Therefore, the dip model predicts an almost pure
cosmic-ray composition with small spread in primary masses.

In a recent publication, the distributions of depths of shower maximum
\xmaxs\ (the atmospheric depth where the number of particles in the
air shower reaches a maximum value) observed at the Pierre Auger
Observatory were interpreted in terms of primary
masses~\cite{longXmax_fits2014} based on current hadronic interaction
models.  The results suggest a mixed mass composition, but there are
differences between the interaction models, and a clear rejection of
the dip model is hindered due to the uncertainties in modeling
hadronic interactions\footnote{For indirect tests of the dip model
  using cosmogenic neutrinos, see e.g.~\cite{dip_neutrinos2015} and
  references therein.}.  Specifically, around the ankle, a very light
composition consisting of proton and helium nuclei only is favoured
using QGSJetII-04~\cite{qgsjetii04} and Sibyll~2.1~\cite{sibyll21},
while for \eposlhc{}~\cite{eposlhc}, intermediate nuclei (of mass
number $A \simeq 14$) contribute. The spread of masses in the primary
beam near the ankle, estimated from the moments of the
\xmaxs\ distributions measured at the Pierre Auger
Observatory~\cite{PAO_Xmax_JCAP2013,longXmax2014}, depends as well on
the details of the hadronic interactions and the results include the
possibility of a pure mass composition.  Observations of \xmaxs\ by
the Telescope Array in the northern hemisphere were found compatible
within uncertainties to both a pure proton
composition~\cite{TA_MD2014} and to the data from the Auger
Observatory~\cite{WGmass_UHECR2014}.

In this report, by exploiting the correlation between two observables
registered simultaneously with different detector systems, we present
results on the spread of primary masses in the energy range
\logenr{18.5}{19.0}, i.e. around the ankle feature. These results are
robust with respect to experimental systematic uncertainties and to
the uncertainties in the description of hadronic interactions.

%%%%%%%%%%%%%%%%%%%%%%%%%%%
%% Method and observables.
%%%%%%%%%%%%%%%%%%%%%%%%%%%

\section{Method and observables}

We follow~\cite{Risse_corrAP2012}
where it was proposed to exploit the correlation between \xmaxs\ and
the number of muons $N_\mu$ in air showers
%, measured independently by two detector systems,
to determine whether the mass composition is pure or mixed.  The
measurement must be performed by two independent detector systems to
avoid correlated detector systematics.  For pure cosmic-ray mass
compositions, correlation coefficients close to or larger than zero
are found in simulations.  In contrast, mixed mass compositions show a
negative correlation, which can be understood as a general
characteristic of air showers well reproduced within a semi-empirical
model~\cite{matthews_heitler}: heavier primaries have on average a
smaller \xmaxs\ ($\Delta\xmaxs\sim-\Delta\ln A$) and larger $N_\mu$
($N_\mu\sim A^{1-\beta}$, $\beta\simeq0.9$~\cite{alvarez2002}), such
that for mixtures of different primary masses, a negative correlation
appears.
% from the simultaneous measurement of \xmaxs\ and $N_\mu$. 
This way, the correlation coefficient can be used to determine
the spread $\sigma(\ln A)$ of primary masses, given by
$\sigma(\ln A)=\sqrt{\langle\ln^2 A\rangle-\langle\ln A\rangle^2}$
where $\langle\ln A\rangle=\sum_i f_i\ln A_i$ and
$\langle\ln^2 A\rangle=\sum_i f_i\ln^2 A_i$
with $f_i$ being the relative fraction of mass $A_i$.
In particular, a more negative correlation indicates a larger
spread of primary masses.

At the Pierre Auger Observatory, the fluorescence telescopes allow a
direct measurement of \xmaxs\ and energy, and the surface array of
water Cherenkov detectors provide a significant sensitivity to muons:
for zenith angles between 20 and 60 degrees, muons contribute about
40\% to 90\%~\cite{auger_icrc2013_kegl} of \smill, the total signal at
a core distance of 1000 m.  Due to this unique feature the proposed
method can be adapted via replacement of $N_\mu$ by \smill, which is a
fundamental observable of the surface array.

Since \smill\ and \xmaxs\ of an air shower depend on its energy
and, in case of \smill, also on its zenith angle, \smill\ and
\xmaxs\ are scaled to a reference energy and zenith angle. This way we
avoid a decorrelation between the observables from combining different
energies and zenith angles in the data set. \smill\ is scaled to
38$^\circ$ and 10~EeV using the parameterisations
from~\cite{schulz_spec_icrc13}. \xmaxs\ is scaled to 10~EeV using an
elongation rate $d\langle\xmaxs\rangle/d\lg(E/{\rm
  eV})=58$~\gsm/decade, an average value with little variation between
different primaries and interaction models~\cite{longXmax2014}. Here,
these scaled quantities will be denoted as \xmax\ and \smilla.  Thus,
\xmax\ and \smilla\ are the values of \xmaxs\ and \smill\ one would
have observed, had the shower arrived at 38$^\circ$ and 10~EeV. It
should be noted that the specific choice of the reference values is
irrelevant, since a transformation to another reference value shifts
the data set as a whole, leaving the correlation coefficient
invariant.

As a measure of the correlation between \xmax\ and \smilla\ the
ranking coefficient \rGsmilla\ introduced by Gideon and
Hollister~\cite{Gideon_JASA87} is taken. Conclusions are unchanged
when using other definitions of correlation coefficients, including
the coefficients of Pearson or Spearman, or other
ones~\cite{Niven_corr2012}. As for any ranking coefficient, the
\rGs\ value is invariant against any modifications leaving the ranks
of events unchanged (in particular to systematic shifts in the
observables). The main distinction from other ranking coefficients is
that the values of ranks are not used directly to calculate
\rGs. Rather the general statistical dependence between \xmax\ and
\smilla\ is estimated by counting the difference in numbers of events
with ranks deviating from the expectations for perfect correlation and
anti-correlation. Thus, the contribution of each event is equal to 0
or 1, making \rGs\ less sensitive to a removal of individual events,
as it will be discussed also below.

The dependence of the statistical uncertainty $\Delta \rGs$ on the
number of events $n$ in a set and on the \rGs\ value itself was
determined by drawing random subsamples from large sets of simulated
events with different compositions. The statistical uncertainty can be
approximated by $\Delta \rGs \simeq 0.9/\sqrt{n}$. For the event set
used here $\Delta \rGs({\rm data})=0.024$.

%%%%%%%%%%%%%%%%%%%%%%%%
%% Data and simulations.
%%%%%%%%%%%%%%%%%%%%%%%%

\section{Data and simulations}

The analysis is based on the same hybrid events as
in~\cite{longXmax2014} recorded by both the fluorescence and the
surface detectors during the time period from 01.12.2004 until
31.12.2012. The data selection procedure, described in detail
in~\cite{longXmax2014}, guarantees that only high-quality events are
included in the analysis and that the mass composition of the selected
sample is unbiased. The reliable reconstruction of \smill\ requires an
additional application of the fiducial trigger cut (the station with
the highest signal should have at least 5 active neighbour
stations). This requirement does not introduce a mass composition
bias since in the energy and zenith ranges considered the surface
detector is fully efficient to hadronic
primaries~\cite{auger_SDtrigger2010,ltp_auger2011}.  Selecting
energies of \logenr{18.5}{19.0} and zenith angles $<$65$^\circ$, the
final data set contains 1376 events.  The resolution and systematic
uncertainties are about 8\% and 14\% in primary
energy~\cite{auger_icrc2013_verzi}, $<20$~${\rm g~cm^{-2}}$ and
10~${\rm g~cm^{-2}}$ in \xmaxs~\cite{longXmax2014}, and $<12$\% and
5\%~\cite{aveSD_icrc2007} in \smill, respectively.

% The Monte-Carlo (MC)
The simulations were performed with CORSIKA~\cite{corsika}, using
\eposlhc{}, QGSJetII-04 or Sibyll~2.1 as the high-energy hadronic
interaction model, and FLUKA~\cite{fluka2008} as the low-energy model.
All events passed the full detector simulation and
reconstruction~\cite{offline07} with the same cuts as applied to data.
For each of the interaction models the shower library contains at
least 10000 showers for proton primaries and $5000-10000$ showers each
for helium, oxygen and iron nuclei.

%%%%%%%%%%%
%% Results.
%%%%%%%%%%%

\section{Results}

\begin{figure}[hbt]
\includegraphics[width=0.5\textwidth]{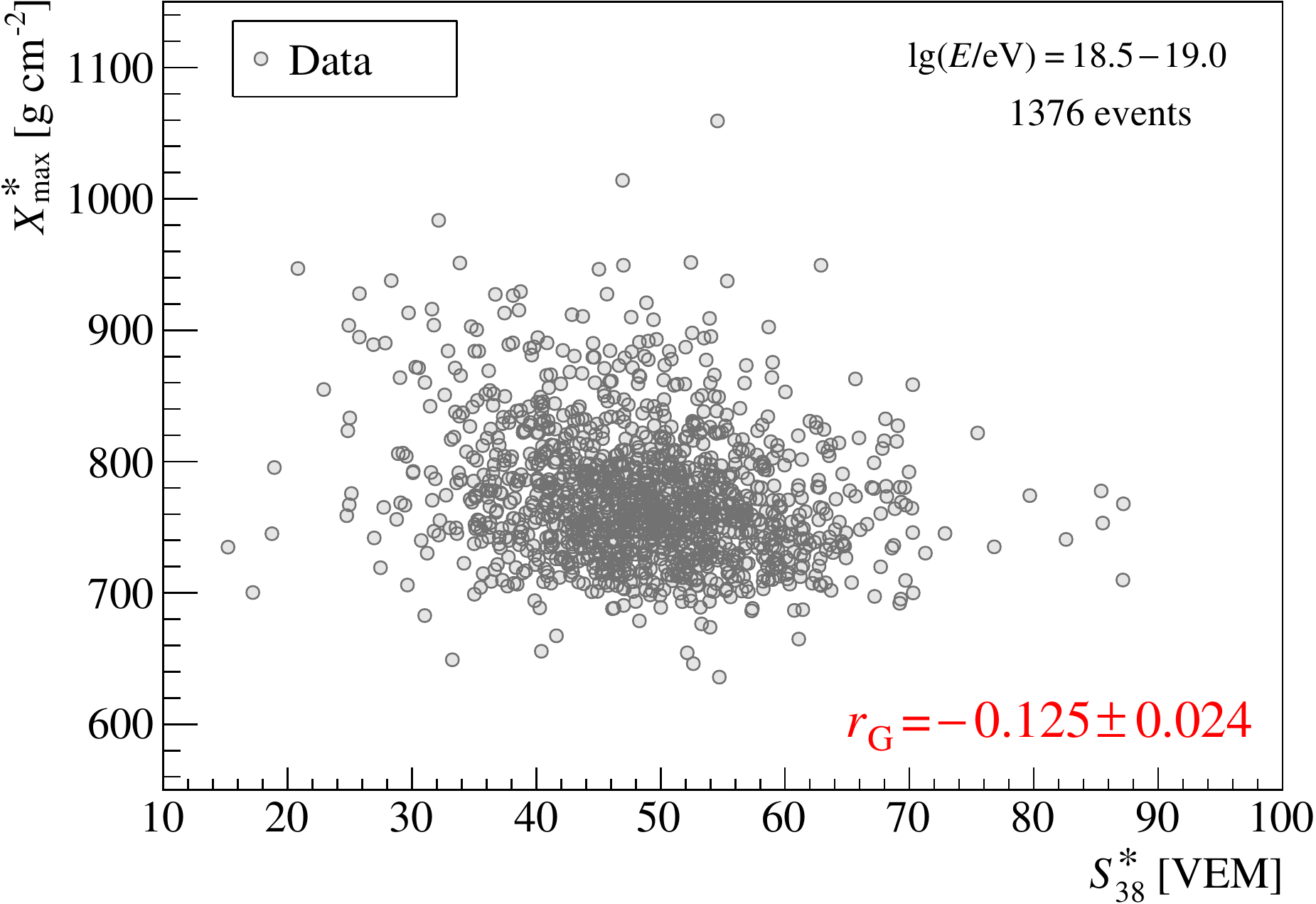}
\includegraphics[width=0.5\textwidth]{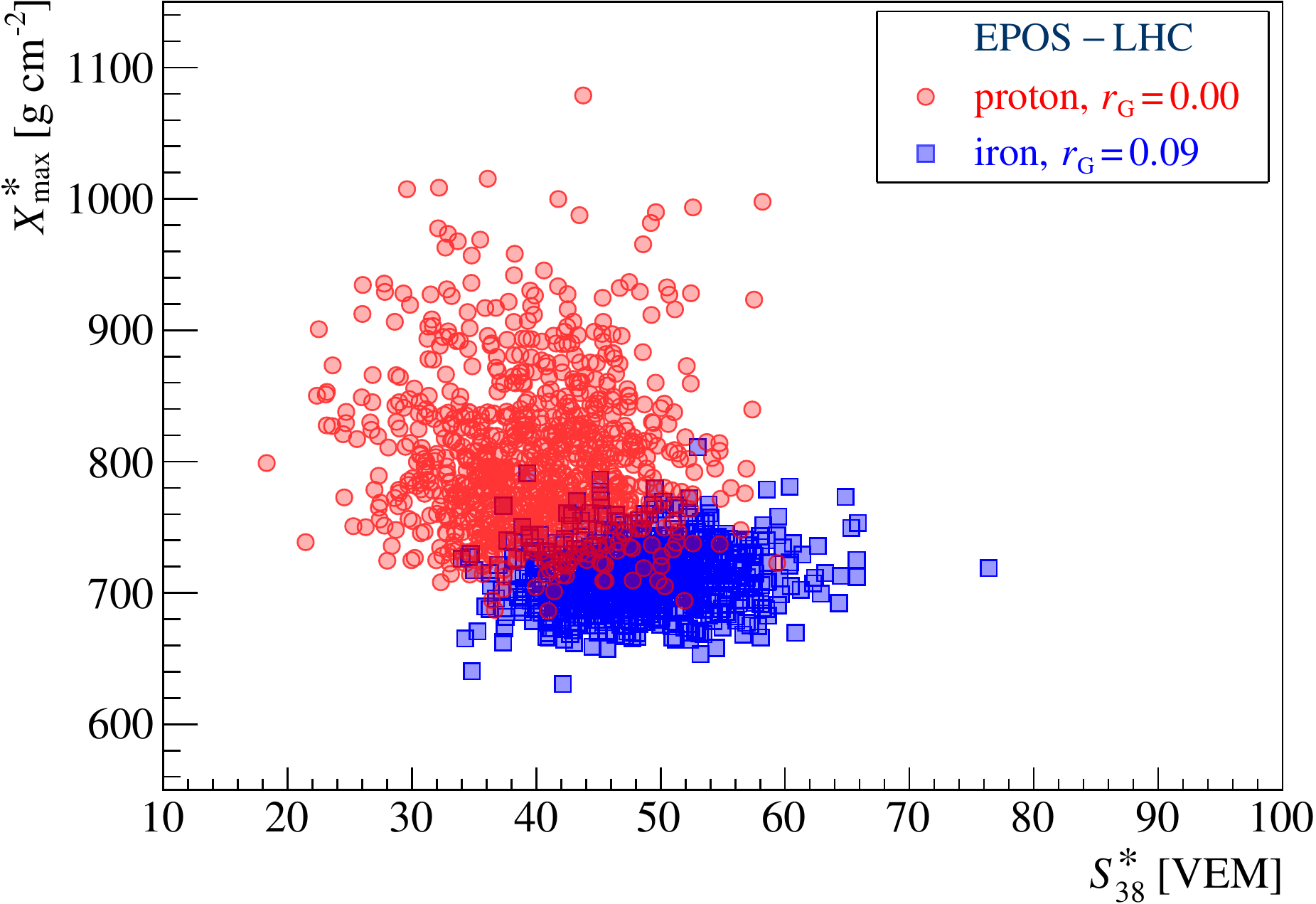}\\[0.2cm]
\caption{Left: measured \xmax\ {\it vs.} \smilla\ for
  \logenr{18.5}{19.0}.  Right: the same distribution for 1000 proton and
  1000 iron showers simulated with \eposlhc{}.}
\label{fig:scatter}
\end{figure}

The observed values of \xmax\ {\it vs.}  \smilla\, are displayed in
Fig.~\ref{fig:scatter}.  As an illustration, proton and iron
simulations for \eposlhc{} are shown as well, but one should keep in
mind that in this analysis we do not aim at a direct comparison of
data and simulations in terms of absolute values. In contrast to the
correlation analysis such a comparison needs to account for
systematics in both observables and suffers from larger uncertainties
from modeling of hadronic interactions.

%% ------ Figs.and Tab. ---------------

\begin{table}[hbt]
\caption{Observed \rGsmilla\ with statistical uncertainty, and
  simulated \rGsmilla\ for various compositions using different
  interaction models (statistical uncertainties are $\approx$0.01).}
\label{tab:rsmilla065g}
\begin{center}
\renewcommand{\tabcolsep}{9pt}
\begin{tabular}{cccc}
  \hline
  data&\multicolumn{3}{c}{$-0.125\pm0.024\,(\mathrm{stat})$}\\
  \hline
  \hline
  &EPOS-LHC & QGSJetII-04 & Sibyll 2.1 \\
  \hline
  $p$&0.00&0.08&0.06\\
  He&0.10&0.16&0.14\\
  O&0.09&0.16&0.17\\
  Fe&0.09&0.13&0.12\\[0.1cm]
$0.5\,p\,-\,0.5\,\mathrm{Fe}$&-0.37&-0.32&-0.31\\[0.1cm]
$0.8\,p\,-\,0.2\,\mathrm{He}$&0.00&0.07&0.05\\
  \hline
\end{tabular}
\end{center}

\end{table}

\begin{figure}[!t]
\includegraphics[width=0.5\textwidth]{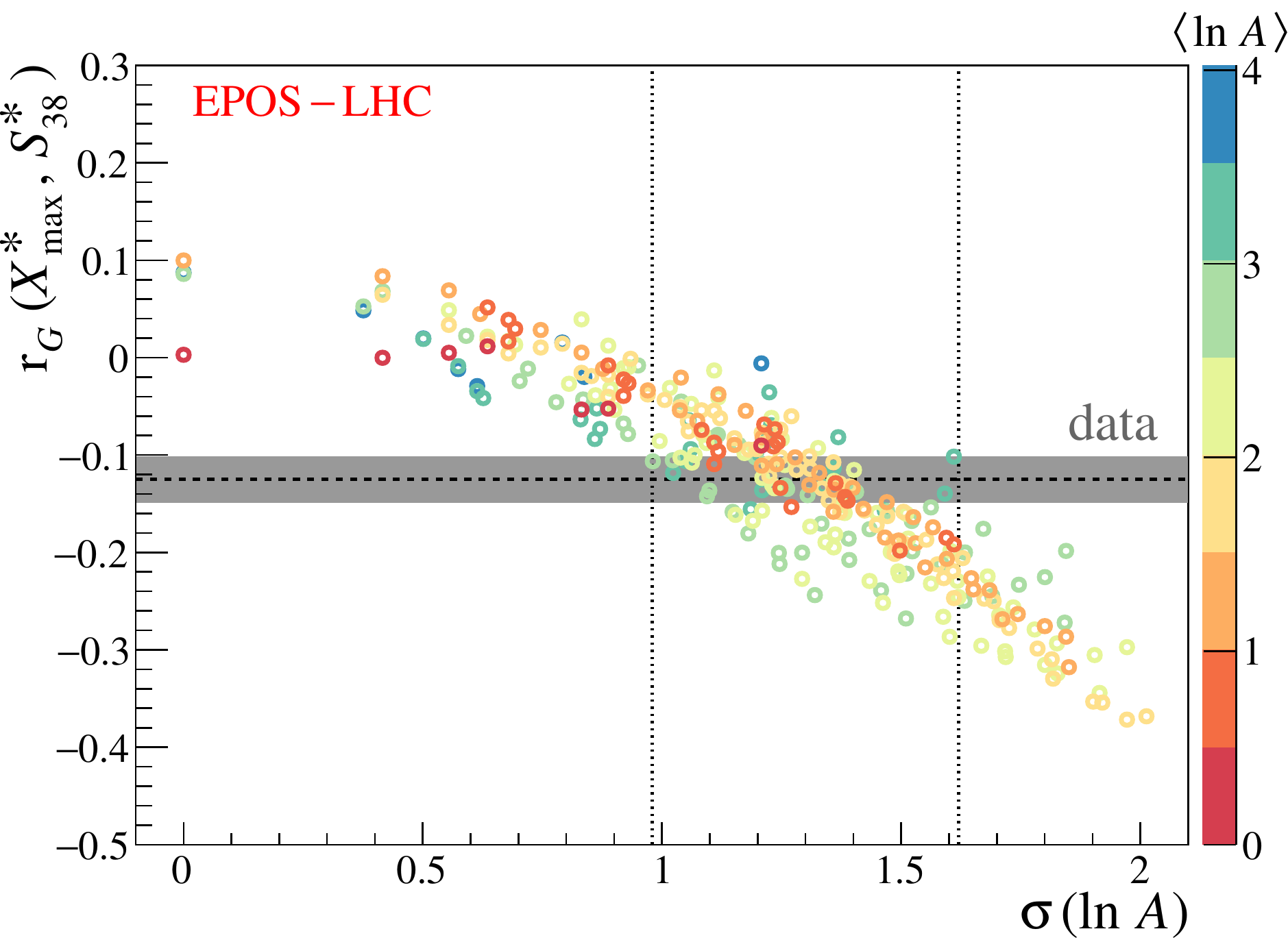}
\includegraphics[width=0.5\textwidth]{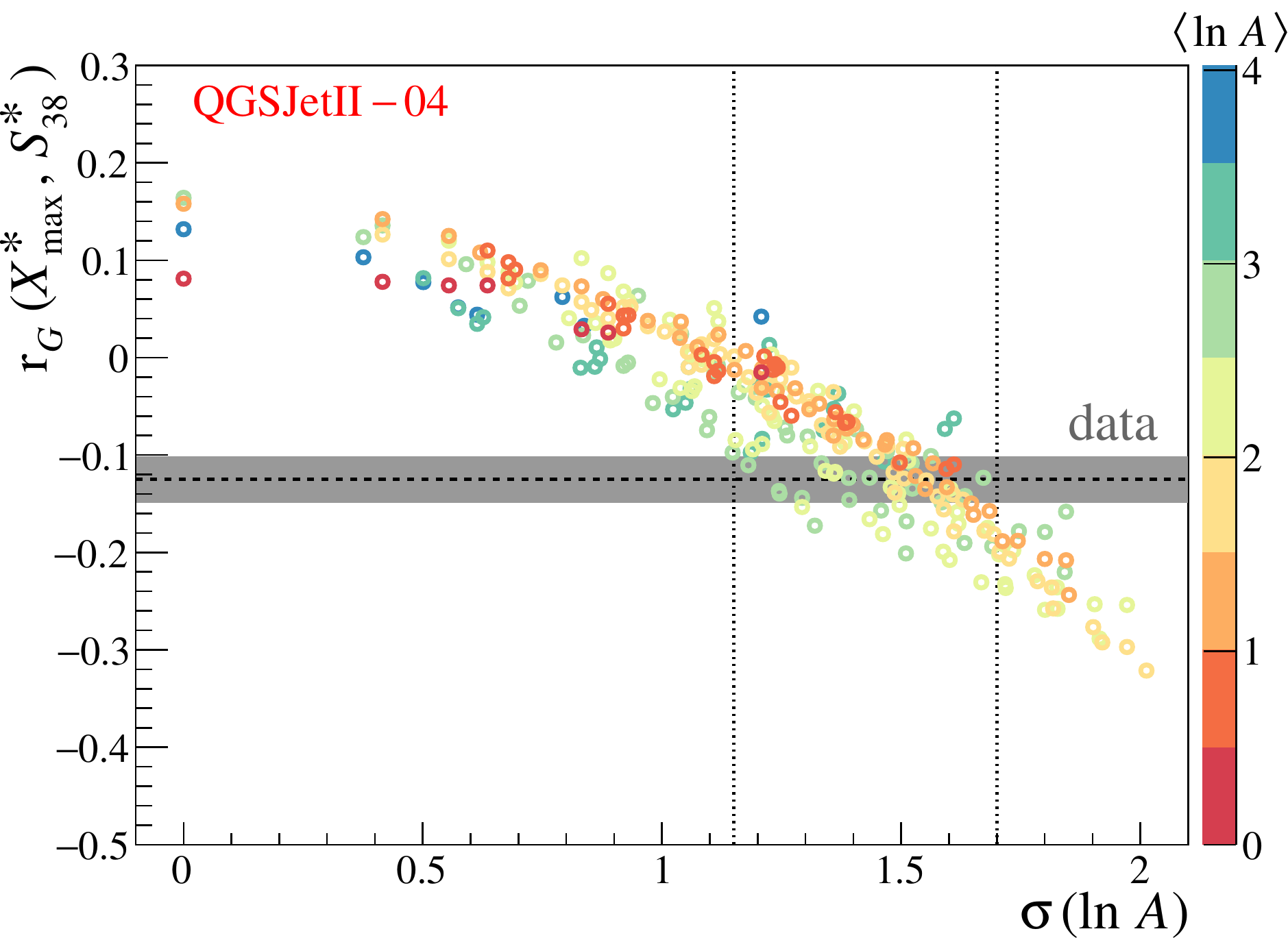}
\caption{Dependence of the correlation coefficients \rGs\ on
  $\sigma(\lnA)$ for \eposlhc{} (left) and QGSJetII-04 (right). Each
  simulated point corresponds to a mixture with different fractions of
  protons, helium, oxygen and iron nuclei, the relative fractions
  changing in 0.1 steps (4 points for pure compositions are grouped at
  $\sigma(\lnA) = 0$). Colors of the points indicate \meanlnA\ of the
  corresponding simulated mixture. The shaded area shows the observed
  value for the data. Vertical dotted lines indicate the range of
  $\sigma(\lnA)$ in simulations compatible with the observed
  correlation in the data.}
\label{fig:xmaxs_vs_rmslna}
\end{figure}

\begin{figure}[!t]
\begin{center}
\includegraphics[width=0.7\textwidth]{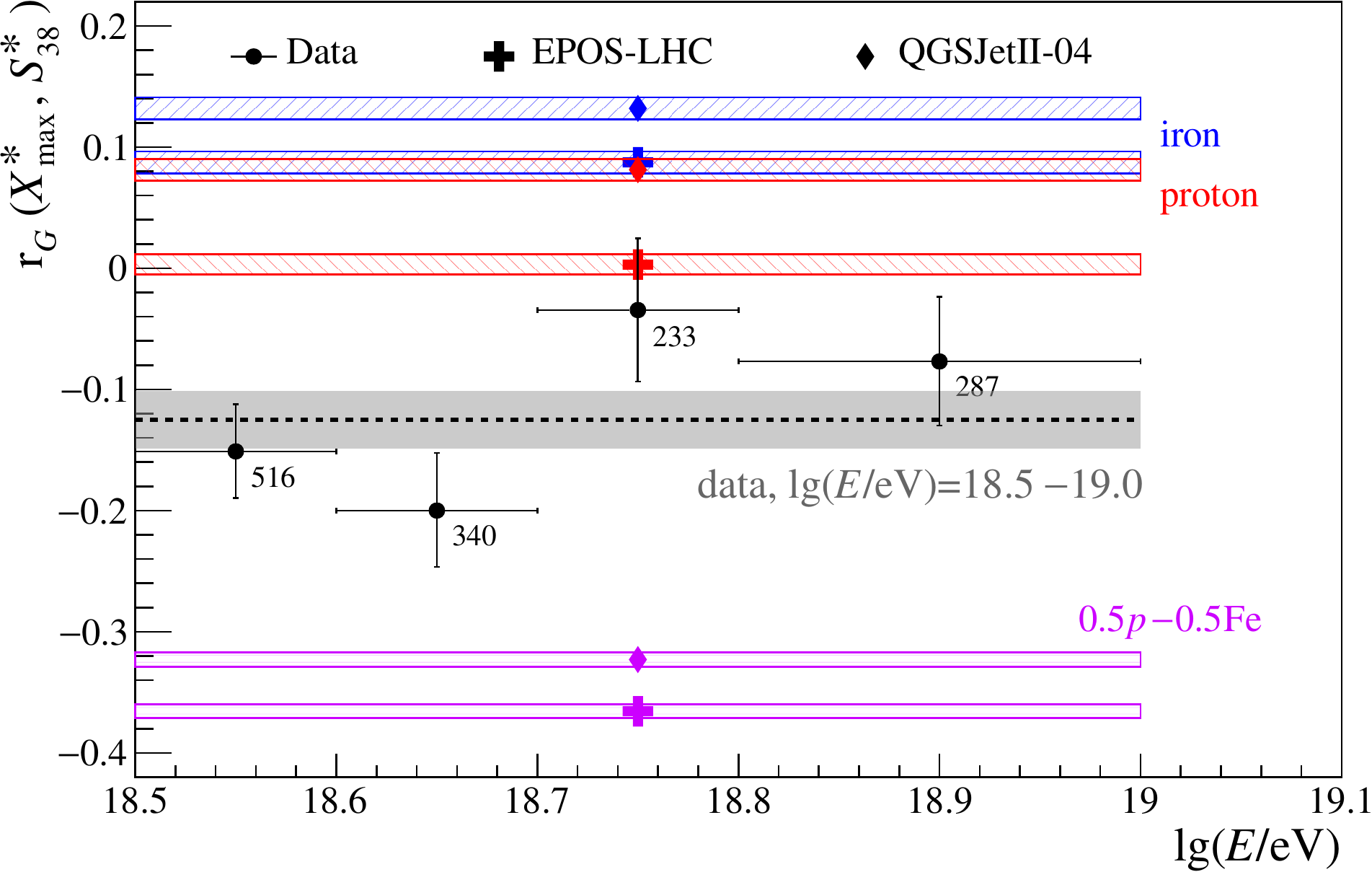}
\end{center}
\caption{The correlation coefficients \rGs\ for data in the energy
  bins
  $\lg(E/\mathrm{eV})=18.5-18.6;\,18.6-18.7;\,18.7-18.8;\,18.8-19.0$. Numbers
  of events in each bin are given next to the data points. The gray
  band shows the measured value for data in the whole range
  \logenr{18.5}{19.0}. Predictions for the correlations \rGs\ in this
  range for pure proton and iron compositions, and for the extreme mix
  $0.5\,p\,-\,0.5\,\mathrm{Fe}$ from \eposlhc{} and QGSJetII-04 are
  shown as hatched bands (for Sibyll~2.1 values are similar to those
  of QGSJetII-04). The widths of the bands correspond to statistical
  errors.}
\label{fig:energybins}
\end{figure}
%% -----------------------

In Table~\ref{tab:rsmilla065g}, the observed \rGsmilla\ is given along
with simulated \rGs\ values for pure compositions
($\sigma(\lnA)=0$) and for the maximum spread of masses
$0.5\,p\,-\,0.5\,\mathrm{Fe}$ ($\sigma(\lnA)\simeq2$) for all three
interaction models.  For the data, a negative correlation
of $\rGsmilla=-0.125\pm0.024\,(\mathrm{stat})$ is found.  For proton
simulations correlations are close to zero or positive in all
models. Pure compositions of heavier primaries show even more positive
correlations ($\rGs\geq0.09$) than for protons.  Hence, observations
cannot be reproduced by any pure composition of mass $A \ge 1$,
irrespective of the interaction model chosen.

In the proton dip model, even small admixtures of heavier nuclei, such
as a $15-20\%$ helium fraction at the sources, were shown to upset the
agreement of the pair-production dip of protons with the observed
flux~\cite{berezinsky_dip_prl2005,berezinsky_dip_ankle_2006,wibig_dip2005,allard_dip2005}. The
values of \rGs\ in simulations for a mixture at Earth of
$0.8\,p\,-\,0.2\,\mathrm{He}$ are added in
Table~\ref{tab:rsmilla065g}. They are essentially unaltered compared
to the pure proton case and equally inconsistent to the observed
correlation.

Further, the correlation is found to be non-negative
$\rGsmilla\gtrsim0$ for all $p\,-\,\mathrm{He}$ mixtures. Thus, the
presence of primary nuclei heavier than helium $A>4$ is required to
explain the data.

We also checked the case of $\mathrm{O}\,-\,\mathrm{Fe}$ mixtures,
i.e. a complete absence of light primaries. A minimum value of
$\rGs\approx-0.04$ is reached for mixtures produced with \eposlhc{}
for fractions close to $0.5\,\mathrm{O}\,-\,0.5\,\mathrm{Fe}$. With
smaller significance, light primaries therefore appear required as
well to describe the observed correlation.

In Fig.~\ref{fig:xmaxs_vs_rmslna} the dependence of the simulated
correlation \rGsmilla\ on the spread $\sigma(\lnA)$ is shown for
\eposlhc{} and QGSJetII-04 (results for Sibyll~2.1 are almost
identical to those of QGSJetII-04). A comparison with the data
indicates a significant degree of mixing of primary masses.
Specifically, $\sigma(\lnA)\simeq 1.35 \pm 0.35$, with values of
$\sigma(\lnA)\simeq 1.1-1.6$ being consistent with expectations from
all three models. The fact that differences between models are
moderate reflects the relative insensitivity of this analysis to
details of the hadronic interactions.

In Fig.~\ref{fig:energybins} the observed values of \rGs\ are
presented in four individual energy bins. From simulations, only a
minor change of \rGs\ with energy is expected for a constant
composition.  The data are consistent with a constant \rGs\ with
$\chi^2 / {\rm dof} \simeq 6.1/3$ ($P\simeq 11\%$).  Allowing for an
energy dependence, a straight-line fit gives a positive slope and
$\chi^2 / {\rm dof} \simeq 3.2/2$ ($P\simeq 20\%$).  More data are
needed to determine whether a trend towards larger \rGs\ (smaller
$\sigma(\lnA)$) with energy can be confirmed.

\section{Uncertainties}

\subsection{Cross-checks}

Several cross-checks were performed.  In all cases, the conclusions
were found to be unchanged. The cross-checks included: (i) a division
of the data set in terms of time periods, FD telescopes or zenith
angle ranges; (ii) variations of the event selection criteria; (iii)
variations of the scaling functions when transforming to the reference
zenith angle and energy; (iv) adopting other methods to calculate the
correlation coefficient~\cite{Niven_corr2012}; and (v) studying the
effect of possible `outlier' events.  Regarding (iv), the smallest
difference between the data and pure compositions is found for
\eposlhc{} protons and it is $5.2\sigma_{\rm stat}$ for \rGs\
(c.f. Table~\ref{tab:rsmilla065g}), and $\ge7\sigma_{\rm stat}$ for
Pearson and Spearman correlation coefficients. As an example of the last point
(v), events were artificially removed from the data set so as to
increase the resulting value of \rGs\ as much as possible, i.e., to
bring it closer to the predictions for pure compositions.  Removing 20
events in this way increased the value of \rGs\ by $\sim0.01$ only.
For removals of sets of 100 arbitrary events, the maximum increase was
$\sim0.02$.  This robustness of \rGs\ against the influence of
individual events and even sub-groups of events was a main reason for
choosing it in this analysis.

%%%%%%%%%%%%%%%%%%%%%%%%%%%
%% Systematics.
%%%%%%%%%%%%%%%%%%%%%%%%%%%%

\subsection{Systematic uncertainties}

Due to the analysis method and the choice of using a correlation
coefficient, systematics are expected to play only a minor role (for
the special case of hadronic uncertainties see below): separate
systematics in the observables \xmaxs\ and \smill\ have no effect on
\rGs, and the measurement of the two observables by independent
detectors avoids correlated systematics. Even a correlated systematic
leaves \rGs\ invariant as long as the ranks of the events are
unchanged.  Also if there were a more subtle issue affecting the ranks
of the observed events that might have gone unnoticed so far and could
require future correction (e.g.\ updated detector calibrations or
atmospheric parameters affecting only part of the data), we note that
this typically leads to a de-correlation of the uncorrected data set,
i.e., to an underestimation of the present value of $|\rGs|$.
Moreover, the main conclusion about the spread of primary masses
results from the {\it difference} between data and simulations which
remains robust for anything affecting the two in a similar way such
as, for instance, during reconstruction.

As an illustration, new data sets were created from the observed one
by artificially introducing energy and zenith angle dependent `biases'
in \xmax\ (up to 10~\gsm) and \smilla\ (up to 10\%) (it should be
stressed that these are arbitrary modifications).  The values of \rGs\
changed by $\lesssim0.01$, which is well below the statistical
uncertainty.  A value of 0.01 is taken as a conservative estimate of
the systematic uncertainty.

The systematics in energy affect the energy bin that the observed
spread is assigned to, which may be shifted by $\pm14\%$.  The
difference between simulation and data is left invariant since
\rGs\ is practically constant with energy for a given composition.

%%%%%%%%%%%%%%%%%%%%%%%%%%%
%% Hadronics.
%%%%%%%%%%%%%%%%%%%%%%%%%%%%

\subsection{Uncertainties in hadronic interactions}

Current model predictions do not necessarily bracket the correct
shower behaviour. In fact, measurements of the muon content from the
Auger Observatory indicate a possible underestimation of muons in
simulations~\cite{auger_icrc2013_farrar,has_muons2015}.  Therefore we
studied whether adjustments of hadronic parameters in simulations
could bring primary proton predictions into full agreement with the
data. The focus is on protons since heavier nuclei, due to the
superposition of several nucleons and the smaller energy per nucleon,
would require even larger adjustments.

Firstly, the (outdated) pre-LHC versions of EPOS and
QGSJetII were checked.
Despite the updates, values of \rGs\ differ by less
than $0.02$ from the current versions.

Secondly, an {\it ad-hoc} scaling of shower muons was applied in
simulations. Different approaches were tested: a constant increase of
the muon number; a zenith-angle dependent increase; and an
accompanying increase of the electromagnetic component as motivated
from shower universality~\cite{univer_rio_2013_talk}.  For an
effective muon scaling by a factor~$\simeq1.3$ as suggested by
data~\cite{auger_icrc2013_farrar,has_muons2015} the simulated
\rGs\ values were reduced by $\lesssim0.03$.  While possibly slightly
decreasing the difference with the data, such a shift is insufficient
to match expectations for pure compositions with data.

Thirdly, following the approach described
in~\cite{ulrich_hmulti_prd2011} and using CONEX~\cite{conex_2006} with
the 3D option for an approximate estimation of the ground signal, the
effect on \rGs\ was studied when modifying some key hadronic
parameters in the shower simulations.  Increasing separately the
cross-section, multiplicity, elasticity, and pion charge ratio by a
factor growing linearly with $\lg E$ from 1.0 at $10^{15}$~eV to 1.5
at $10^{19}$~eV compared to the nominal values ($f_{19}=1.5$,
cf.~\cite{ulrich_hmulti_prd2011}), \rGs\ turned out to be essentially
unaffected except for the modified cross-section where the value was
decreased by $\Delta\rGs\approx-0.06$.  Despite the large increase of
the cross-section assumed, this shift is still insufficient to explain
the observed correlation.  Moreover, $\Delta$\rGs\ shows in this case
a strong dependence on zenith angle ($\simeq 0.0$ for 0$-$45$^\circ$
and $\simeq -0.1$ for 45$-$60$^\circ$) making the predictions
inconsistent with the data.  It should be noted that any such
modification is additionally constrained by other data of the Auger
Observatory such as the observed
\xmaxs\ distributions~\cite{longXmax2014} and the proton-air
cross-section at $\lg(E/{\rm
  eV})\simeq18.25$~\cite{prl2012_cross,ulrich_icrc2015}.

%%%%%%%%%%%
%% Discussion.
%%%%%%%%%%%

\section{Discussion}

A negative correlation of $\rGsmilla = -0.125\pm0.024\,({\rm stat})$
is observed.  Simulations for any pure composition with \eposlhc{},
QGSJetII-04 and Si\-byll~2.1 give $\rGs\ge 0.00$ and are in conflict
with the data. Equally, simulations for all proton\,--\,helium
mixtures yield $\rGs\ge 0.00$. The observations are naturally
explained by a mixed composition including nuclei heavier than helium
$A>4$, with a spread of masses $\sigma(\lnA)\simeq 1.35 \pm 0.35$.

Increasing artificially the muon component or changing some key
hadronic parameters in shower simulations leaves the findings
essentially unchanged.  Thus, even with regard to hadronic interaction
uncertainties, a scenario of a pure composition is implausible as an
explanation of our observations.  Possible future attempts in that
direction may require fairly exotic solutions. In any case, they are
highly constrained by the observations presented here as well as by
previous Auger results.

The minor dependence of the mass spread determined in this analysis
from hadronic uncertainties allows one to test the self-consistency of
hadronic interaction models when deriving the composition from other
methods or observables
(e.g.~\cite{longXmax2014,longXmax_fits2014,augerMPD2014,auger_azasy2016}).
As mentioned in the beginning, when interpreting the
\xmaxs\ distributions alone in terms of fractions of
nuclei~\cite{longXmax_fits2014}, different results are found depending
on the model: using QGSJetII-04 or Sibyll~2.1, one infers values of
$\sigma(\ln A)\approx0.7$ and would expect $\rGs\approx0.08$.  This is
at odds with the observed correlation and indicates shortcomings in
these two models.  Using \eposlhc{}, values of $\sigma(\ln
A)\approx1.2$ and $\rGs\approx-0.094$ are obtained, in better
agreement with the observed correlation.

The conclusion that the mass composition at the ankle is not pure but
instead mixed has important consequences for theoretical source
models.  Proposals of almost pure compositions, such as the dip
scenario, are disfavoured as the sole explanation of ultrahigh-energy
cosmic rays.  Along with the previous Auger
results~\cite{longXmax_fits2014,PAO_Xmax_JCAP2013,longXmax2014}, our
findings indicate that various nuclei, including masses $A > 4$, are
accelerated to ultrahigh energies ($>10^{18.5}$~eV) and are able to
escape the source environment.

% created on 2016-06-15

\section*{Acknowledgments}

\begin{sloppypar}
The successful installation, commissioning, and operation of the Pierre Auger Observatory would not have been possible without the strong commitment and effort from the technical and administrative staff in Malarg\"ue. We are very grateful to the following agencies and organizations for financial support:
\end{sloppypar}

\begin{sloppypar}
Comisi\'on Nacional de Energ\'\i{}a At\'omica, Agencia Nacional de Promoci\'on Cient\'\i{}fica y Tecnol\'ogica (ANPCyT), Consejo Nacional de Investigaciones Cient\'\i{}ficas y T\'ecnicas (CONICET), Gobierno de la Provincia de Mendoza, Municipalidad de Malarg\"ue, NDM Holdings and Valle Las Le\~nas, in gratitude for their continuing cooperation over land access, Argentina; the Australian Research Council; Conselho Nacional de Desenvolvimento Cient\'\i{}fico e Tecnol\'ogico (CNPq), Financiadora de Estudos e Projetos (FINEP), Funda\c{c}\~ao de Amparo \`a Pesquisa do Estado de Rio de Janeiro (FAPERJ), S\~ao Paulo Research Foundation (FAPESP) Grants No.\ 2010/07359-6 and No.\ 1999/05404-3, Minist\'erio de Ci\^encia e Tecnologia (MCT), Brazil; Grant No.\ MSMT CR LG15014, LO1305 and LM2015038 and the Czech Science Foundation Grant No.\ 14-17501S, Czech Republic; Centre de Calcul IN2P3/CNRS, Centre National de la Recherche Scientifique (CNRS), Conseil R\'egional Ile-de-France, D\'epartement Physique Nucl\'eaire et Corpusculaire (PNC-IN2P3/CNRS), D\'epartement Sciences de l'Univers (SDU-INSU/CNRS), Institut Lagrange de Paris (ILP) Grant No.\ LABEX ANR-10-LABX-63, within the Investissements d'Avenir Programme Grant No.\ ANR-11-IDEX-0004-02, France; Bundesministerium f\"ur Bildung und Forschung (BMBF), Deutsche Forschungsgemeinschaft (DFG), Finanzministerium Baden-W\"urttemberg, Helmholtz Alliance for Astroparticle Physics (HAP), Helmholtz-Gemeinschaft Deutscher Forschungszentren (HGF), Ministerium f\"ur Wissenschaft und Forschung, Nordrhein Westfalen, Ministerium f\"ur Wissenschaft, Forschung und Kunst, Baden-W\"urttemberg, Germany; Istituto Nazionale di Fisica Nucleare (INFN),Istituto Nazionale di Astrofisica (INAF), Ministero dell'Istruzione, dell'Universit\'a e della Ricerca (MIUR), Gran Sasso Center for Astroparticle Physics (CFA), CETEMPS Center of Excellence, Ministero degli Affari Esteri (MAE), Italy; Consejo Nacional de Ciencia y Tecnolog\'\i{}a (CONACYT) No.\ 167733, Mexico; Universidad Nacional Aut\'onoma de M\'exico (UNAM), PAPIIT DGAPA-UNAM, Mexico; Ministerie van Onderwijs, Cultuur en Wetenschap, Nederlandse Organisatie voor Wetenschappelijk Onderzoek (NWO), Stichting voor Fundamenteel Onderzoek der Materie (FOM), Netherlands; National Centre for Research and Development, Grants No.\ ERA-NET-ASPERA/01/11 and No.\ ERA-NET-ASPERA/02/11, National Science Centre, Grants No.\ 2013/08/M/ST9/00322, No.\ 2013/08/M/ST9/00728 and No.\ HARMONIA 5 -- 2013/10/M/ST9/00062, Poland; Portuguese national funds and FEDER funds within Programa Operacional Factores de Competitividade through Funda\c{c}\~ao para a Ci\^encia e a Tecnologia (COMPETE), Portugal; Romanian Authority for Scientific Research ANCS, CNDI-UEFISCDI partnership projects Grants No.\ 20/2012 and No.194/2012 and PN 16 42 01 02; Slovenian Research Agency, Slovenia; Comunidad de Madrid, Fondo Europeo de Desarrollo Regional (FEDER) funds, Ministerio de Econom\'\i{}a y Competitividad, Xunta de Galicia, European Community 7th Framework Program, Grant No.\ FP7-PEOPLE-2012-IEF-328826, Spain; Science and Technology Facilities Council, United Kingdom; Department of Energy, Contracts No.\ DE-AC02-07CH11359, No.\ DE-FR02-04ER41300, No.\ DE-FG02-99ER41107 and No.\ DE-SC0011689, National Science Foundation, Grant No.\ 0450696, The Grainger Foundation, USA; NAFOSTED, Vietnam; Marie Curie-IRSES/EPLANET, European Particle Physics Latin American Network, European Union 7th Framework Program, Grant No.\ PIRSES-2009-GA-246806; and UNESCO.
\end{sloppypar}

%\input{2016-04-acknowledgments.tex}
%\bibliographystyle{/home/yushkov/bin/bibtex/elsarticle-template/model1a-num-names}
%\bibliography{/home/yushkov/bin/bibtex/my_rus}

\end{document}